\documentclass[11pt]{article} 
\usepackage[utf8]{inputenc}

\usepackage{amsfonts}
\usepackage{amsmath}
\usepackage{amssymb}
\usepackage{amsthm}
\usepackage{caption}
\usepackage{color}
    \definecolor{darkgreen}{rgb}{0,0.5,0}
    \definecolor{darkblue}{rgb}{0,0,0.6}
    \definecolor{purple}{rgb}{0.4,.2,0.7}
\usepackage[margin = 2.5cm]{geometry}
\usepackage{graphicx}
\usepackage[hyperfootnotes = false, colorlinks = true, linkcolor = darkblue, citecolor = blue,urlcolor  = darkblue]{hyperref}
\usepackage{subcaption}
\usepackage{appendix}
\usepackage{braket}
\usepackage{tikz}

\allowdisplaybreaks[4]

\begin{document}
\begin{titlepage}
	\renewcommand{\thefootnote}{\fnsymbol{footnote}}
\thispagestyle{empty}
\begin{center}
    ~\vspace{5mm}
    
    {\Large \bf {On heat properties of charged AdS black holes in Gauss-Bonnet gravity coupled with nonlinear electrodynamics}}

    \vspace{0.5in}
    
    {\bf Yang Guo${}^{}$\footnote{guoy@mail.nankai.edu.cn} and Yan-Gang Miao${}^{}$\footnote{Corresponding author: miaoyg@nankai.edu.cn}}

   \vspace{0.5in}

   {\em School of Physics, Nankai University, Tianjin 300071, China}
     
    \vspace{0.5in}

\end{center}

\vspace{0.5in}

\begin{abstract}
  
We investigate the heat properties of charged AdS black holes in the Gauss-Bonnet gravity coupled with nonlinear electrodynamics.
We consider the thermodynamics of black holes from the perspective of heat capacity and show that the nonlinear electrodynamics can be helpful to improve the thermodynamic stability of black holes.
We perform a two-dimensional description in order to reproduce the Hawking temperature, which confirms that the Hawking temperature has an intrinsic topological nature and holds for a higher dimensional spherically symmetric spacetime.
We also analyze the Maxwell equal area law and coexistence curve, and find the existence of van der Waals-like phase transitions based on critical exponents.
Moreover, we deal with a charged AdS black hole in the Gauss-Bonnet gravity coupled with nonlinear electrodynamics as a working material to study holographic heat engines and obtain an exact expression for the efficiency of a rectangular engine cycle. We then discuss the effects of nonlinear electrodynamics and Gauss-Bonnet couplings on the rectangular engine cycle and compare the efficiency of this cycle with that of the Carnot cycle.

\end{abstract}

\vspace{1in}


\setcounter{tocdepth}{3}


\end{titlepage}
\tableofcontents
\renewcommand{\thefootnote}{\arabic{footnote}}
\setcounter{footnote}{0}
\section{Introduction} 
Over the past five decades a remarkable progress has been made in the study of black hole thermodynamics. From the Hawking area theorem~\cite{Hawking:1971vc}, the Bekenstein entropy~\cite{Bekenstein:1973ur} and the four laws of mechanics~\cite{Bardeen:1973gs} to the Hawking radiation in the early 1970s, these discoveries indicate that black holes are no longer simply mechanical systems that can alternatively be described by thermodynamics, but that they are actually thermodynamic systems with temperature and entropy. As a thermodynamic system, a black hole plays an essential role in our understanding of the fundamental relationships among gravity, thermodynamics, and quantum theory. In the framework of an extended phase space, the cosmological constant is regarded~\cite{Caldarelli:1999xj,Kastor:2009wy,Dolan:2010ha} as a thermodynamic variable and identified as the thermodynamic pressure. This has led to a wider field of studies in black hole thermodynamics, where new thermodynamic phase transitions and rich phase structures naturally emerge.

Because the Hawking temperature of astrophysical black holes is~\cite{Carlip:2014pma} far below the temperature of the cosmic microwave background, the Hawking radiation~\cite{Kuang:2017sqa,Gonzalez:2017zdz} has not yet been observed directly. However, the Hawking temperature can be derived in many independent approaches, which gives an indirect evidence that the Hawking radiation actually exists and thus a black hole is a thermodynamic system. In particular, the connection between thermodynamics and topology has recently been established~\cite{Yerra:2022coh,Wei:2021vdx,Zhang:2020kaq,Robson:2018con} for black holes. As a significant step forward, we will reproduce the Hawking temperature by a dimensional truncation, which may be regarded as the progress related~\cite{Ovgun:2019ygw,Robson:2019yzx,Robson:2019wfu} to the topology of spacetime manifold of a black hole.   
       
Although the black hole thermodynamics~\cite{Gregory:2020mmi,Appels:2016uha,Akhmedov:2006pg,Anabalon:2018ydc,Javed:2023iih,Javed:2022bdi} is not a topic of great concerns as it was in the last century, there has been a renewed interest for black hole thermodynamics in recent years that we call the burgeoning field~\cite{Kubiznak:2014zwa} {\em Black Hole Chemistry}. There are a number of novel results in this field showing that black holes behave in many ways quite analogous to common chemical phenomena, such as liquid/gas phase transition and van der Waals fluid behavior~\cite{Poshteh:2013pba,Cai:2013qga,Hennigar:2015wxa,Zeng:2015wtt,Guo:2021wcf,Ovgun:2017bgx}, triple points~\cite{Altamirano:2013uqa,Frassino:2014pha,Wei:2014hba}, Joule-Thomson expansion~\cite{Nam:2019zyk,Assrary:2022uiu,Cisterna:2018jqg} and heat engines~\cite{Belhaj:2015hha,Setare:2015yra,Johnson:2015fva,Chandrasekhar:2016lbd,Johnson:2016pfa,ElMoumni:2021zbp,Ahmed:2019yci}, etc. As is well known, a heat engine works in the way that brings a working substance from a higher  temperature to a lower one. During this process, the heat energy is converted into  the mechanical work. 
The behavior of black hole chemistry naturally allows~\cite{Johnson:2014yja,Johnson:2015ekr} us to treat a black hole as a working substance that is usually regarded as a gas or liquid in thermodynamics. In the field of black hole thermodynamics, the gravity theories with higher derivative curvature terms have received a lot of concerns, in particular, the Lovelock gravity. For the black holes with constant curvature horizon hypersurfaces, their thermodynamic stability is no longer maintained~\cite{Cai:2003kt} and the causality is violated~\cite{Brustein:2017iet} even if these black holes have a positive heat capacity.
In this paper, we deal with the charged AdS black hole in the Gauss-Bonnet gravity coupled with nonlinear electrodynamics as a working material and analyze the related features of corresponding holographic heat engines, such as the efficiency of a rectangular engine cycle, in particular, we investigate the effects on efficiency of heat engines from nonlinear electrodynamics and Gauss-Bonnet couplings.

Our paper is organized as follows. In Sec.~\ref{sec:thermodynamics}, we make a general discussion of black hole thermodynamics from the perspective of heat capacity and investigate the effect of nonlinear electrodynamics on thermodynamic stability.  We then derive in Sec.~\ref{sec:topology} the Hawking temperature in the truncated spacetime by using a topological approach.
In Sec.~\ref{sec:Maxwells}, we investigate Maxwell's equal area law and plot the coexistence curves on the $(P, T)$ plane for the charged AdS black hole in the Gauss-Bonnet gravity coupled with nonlinear electrodynamics. In Sec.~\ref{sec:Exponent}, we compute the critical exponents of this kind of black holes and point out the same behavior as that  of the van der Waals fluid near the critical point. Next, we construct in Sec.~\ref{sec:heat engine} the heat engine and calculate its efficiency, where a charged AdS black hole in the Gauss-Bonnet gravity coupled with  nonlinear electrodynamics is regarded as a working substance.  Finally, we give our conclusions and discussions in Sec.~\ref{sec:con}.
Appendix~\ref{sec:maxwell} is dedicated to the derivation of Maxwell's equal area law.	Throughout the paper we adopt the natural units, $\hbar=c=G=k_{\text{B}}=1$.

\section{Charged AdS black holes in the Gauss-Bonnet gravity coupled with nonlinear electrodynamics and their thermodynamics}
\subsection{Black hole solutions and thermodynamics}
\label{sec:thermodynamics}

 We consider 
 a spherically symmetric metric in the five-dimensional spacetime in the Gauss-Bonnet gravity that includes a negative cosmological constant and couples to nonlinear electrodynamics,
 \begin{eqnarray}
 	ds^2=-f(r)dt^2+\frac{1}{f(r)}dr^2+r^2d\Omega^2_3,
 \end{eqnarray}
 where $d\Omega^2_3$ stands for the standard angular part on $S^3$ and the metric function reads~\cite{Hyun:2019gfz}
\begin{eqnarray}	f(r)=1+\frac{r^2}{4\alpha}\left[1-\sqrt{1-\frac{8\alpha}{l^2}+\frac{8\alpha}{r^4}\left(m+\frac{q^2}{3k}(e^{-k/r^2}-1) \right) } \right],
\end{eqnarray}
where $\alpha$ denotes the Gauss-Bonnet coupling parameter, $l$ the radius of the AdS space,  $m$ the reduced mass, $q$ the magnetic charge, and $k$ the nonlinear electrodynamic parameter. 

Let us analyze the behavior of singularities for this kind black holes. It depends on these parameters. At first, we give the critical value of $m$, $m=\frac{q^2}{3k}$, which is related to $q$ and $k$. Then, we discuss the behavior of singularities in the two regions of the parameter $m$ space, respectively, where one region is  $m<\frac{q^2}{3k}$, and the other $m>\frac{q^2}{3k}$. Now we can turn to the analyses.
	
In the first region, the black hole solution exhibits a branch singularity at $r_s$ determined by
\begin{eqnarray}\label{eq:branch}
1-\frac{8\alpha}{l^2}+\frac{8\alpha }{r_s^4}\left(m+\frac{q^{2}}{3k}(e^{-k/r_s^2}-1)\right)=0.
\end{eqnarray}
The existence of a horizon depends~\cite{Boulware:1985wk} on the sign of Gauss-Bonnet coupling $\alpha$. Because $\alpha$ has been restricted to be non-negative, no event horizons shield the branch singularity, which leads to a naked singularity. It is easy to check  the existence of this naked singularity numerically for an arbitrary charge case.  For the special case of $q\rightarrow0$, we can solve Eq.~\eqref{eq:branch} analytically and give the position at which the branch singularity appears,
\begin{eqnarray}
	r_s=\left( -\frac{8\alpha  l^2 m}{l^2-8 \alpha }\right)^{1/4}, \quad \text{with} \quad 0<\alpha <\frac{l^2}{8},\;\;  m<0.
\end{eqnarray}
For the details, see Refs.~\cite{Wiltshire:1985us,Wiltshire:1988uq}.

In the second region, a curvature singularity exists at the center of this kind of black holes. This point can be seen clearly when we compute the Ricci scalar in the vicinity of $r=0$,
\begin{eqnarray}\label{eq:scacur}
	R=-\frac{5}{\alpha}+\frac{3}{r^2}\sqrt{\frac{2}{\alpha}\left(m-\frac{q^2}{3k}\right)}+\mathcal{O}\left(r^2\right),
\end{eqnarray}
where the curvature singularity is located inside the black hole horizon. However, when the mass equals the critical value, $m=\frac{q^2}{3k}$, the curvature singularity disappears. Thus, the solution is regular because the curvature singularity is replaced by an AdS core,
\begin{eqnarray}
		f(r)\rightarrow 1+\frac{r^2}{l_{\rm eff}^{2}},
\end{eqnarray}
where $l_{\rm eff}$ is the effective radius of AdS space.

 In the extended phase space associated with the cosmological constant $\Lambda$, the thermodynamic pressure is defined as
 \begin{eqnarray}
 	P\equiv-\frac{\Lambda}{8\pi}=\frac{3}{4\pi l^2},
 \end{eqnarray}
  and then the first law of black hole thermodynamics reads,
 \begin{eqnarray}
 	dM=TdS+\Phi dQ+VdP,
 \end{eqnarray}
 where $Q$ stands for the total charge and $\Phi$ for the chemical potential. 
 The AMD mass $M$ of the black hole is identified as enthalpy rather than internal energy, and it has the following relation with the reduced mass $m$,
 \begin{eqnarray}
 	M=\frac{3\pi}{8}m.
 \end{eqnarray} 
 The temperature $T$ and volume $V$ of the black hole are defined as the thermodynamic variables conjugate to  $S$ and $P$, respectively,
 \begin{eqnarray}
 	T\equiv\left( \frac{\partial M}{\partial S}\right)_{P, Q} ,\qquad V\equiv\left( \frac{\partial M}{\partial P}\right)_{S, Q} .
 \end{eqnarray} 
 
\begin{figure}[ht]
	\begin{subfigure}{.5\textwidth}
		\centering
		\includegraphics[width=.9\linewidth]{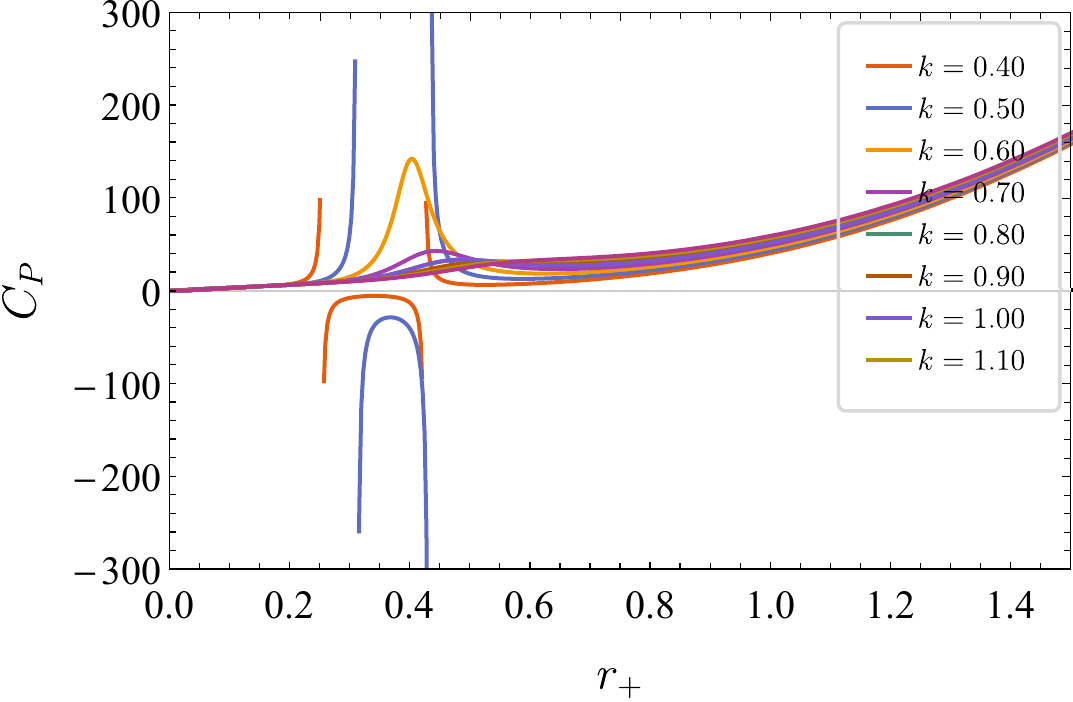}  
		\caption{Gauss-Bonnet coupling parameter $\alpha=0.5$.}
		\label{fig:cp1}
	\end{subfigure}
	\begin{subfigure}{.5\textwidth}
		\centering
		\includegraphics[width=.9\linewidth]{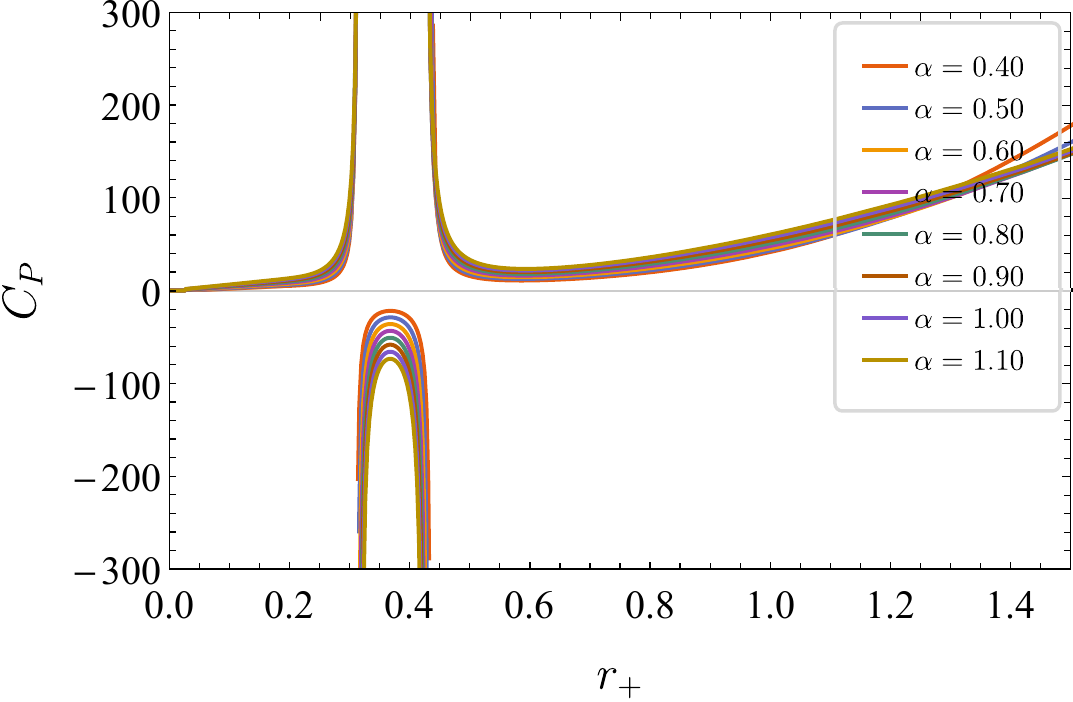}  
		\caption{Nonlinear electrodynamic parameter $k=0.5$.}
		\label{fig:cp2}
	\end{subfigure}
	\caption{The heat capacity with respect to the horizon radius with  $P=0.02$ and $q=0.8$.}
	\label{fig:cp}
\end{figure}

The Hawking temperature $T_{\rm H}$ and entropy $S$ can  be computed,
\begin{eqnarray}
			T_{\rm H}=\frac{f'(r_+)}{4\pi}=\frac{r_+^4 \left(8 \pi  P r_+^2+3\right)-q^2 e^{-{k}/{r_+^2}}}{6 \pi  r_+^3 \left(4 \alpha +r_+^2\right)},\label{eq:temp}
\end{eqnarray}
\begin{eqnarray}
   		S=\int \frac{dM}{T}=\frac{\pi^2r^3_+}{2}\left( 1+\frac{12\alpha}{r^2_+}\right).\label{eq:entropy}
\end{eqnarray}
Here the entropy can also be obtained~\cite{Hyun:2019gfz} via the Wald formula.
Based on Eqs.~(\ref{eq:temp}) and (\ref{eq:entropy}), we calculate the heat capacity  employing the standard thermodynamic method,
 \begin{eqnarray}
 	\label{heatcap}
 	C_P&=&T\left( \frac{\partial S}{\partial T}\right)_P \nonumber\\
 	&=& \frac{3 \pi ^2 r_+^3 \left(4 \alpha +r_+^2\right)^2 \left[r_+^4 \left(8 \pi  P r_+^2+3\right)e^{{k}/{r_+^2}}-q^2\right]}{2 r_+^6  \left[12 \alpha +r_+^2 \left(8 \pi  P \left(12 \alpha +r_+^2\right)-3\right)\right]e^{{k}/{r_+^2}}+2 q^2 \left[-2 k \left(4 \alpha +r_+^2\right)+5 r_+^4+12 \alpha  r_+^2\right]}.
 \end{eqnarray}
As is well known, a black hole system with a positive heat capacity can reach a stable thermal equilibrium with a surrounding heat bath. But a black hole with a negative heat capacity gets~\cite{davies1977thermodynamic} hotter as it radiates energy. Although a negative heat capacity appears to be excluded by classical thermodynamics, it does occur~\cite{lynden1968gravo,thirring1970systems,lynden1999negative} in self-gravitating systems, such as stars and star clusters. In addition, some models with a negative heat capacity are presented~\cite{lakes2008negative,bisquert2005master} in non-equilibrium. In Fig.~\ref{fig:cp1} we can observe that a thermodynamically unstable region with a negative heat capacity appears when $k$ is small ($k=0.40, 0.50$), and that this region disappears when $k$ is  gradually increasing. However, the unstable region does not disappear when we fix the value of $k=0.50$ but vary the value of $\alpha$, see Fig.~\ref{fig:cp2}, indicating that this behavior just depends on the nonlinear electrodynamic parameter $k$. That is, for the spherically symmetric charged AdS black hole in the Gauss-Bonnet gravity, the effect from nonlinear electrodynamics can change an unstable black hole to be a stable one.

 \subsection{Low-dimensional thermodynamic  information storage and topological nature of Hawking radiations}
 \label{sec:topology}
In the algebraic topology, a topological invariance is usually defined as a property preserved under homomorphisms.	And the Euler characteristic is a topological invariant or a number that describes the shape and structure of a manifold. The Euler characteristic $\chi$ of an $n$-dimensional  Riemannian manifold can be defined as follows, 
\begin{eqnarray}
	\chi\equiv\frac{2}{{\rm area}(S^n)}\int_{M^n}^{}\sqrt{g}\,{\cal G}\,d^nx,\label{Chi}
\end{eqnarray}
where area($S^n$) denotes the surface area of an $n$-sphere, and the density $\cal G$ is defined~\cite{morgan1993riemannian} by the Riemann curvature tensor $\mathcal{R}_{\mu_1\mu_2\mu_3\mu_4}$ and the Levi-Civita symbol $\epsilon^{\mu_1\cdots \mu_n}$,
\begin{eqnarray}
	{\cal G}\equiv\frac{1}{2^{n/2}n!g}\epsilon^{\mu_1\cdots \mu_n} \epsilon^{\nu_1\cdots \nu_n}
	\mathcal{R}_{\mu_1\mu_2 \nu_1\nu_2}\cdots \mathcal{R}_{\mu_{n-1} \mu_n \nu_{n-1} \nu_n}.
\end{eqnarray}

Performing a  Wick rotation ($\tau=it$) to the metric of a static and spherically symmetric black hole, we turn the metric from a pseudo-Riemannian quantity to a Riemannian one,
\begin{eqnarray}
	ds^2=f(r)d\tau^2+\frac{1}{f(r)}dr^2+r^2d\Omega^2_{d-2}.
\end{eqnarray}
For example, when $d=4$, the topology of a four-dimensional  Schwarzschild black hole is~\cite{Gibbons291} $\mathbb{R}^2\times S^2$ with the Euler characteristic, $\chi=2$. If we cut off the ($d-2$)-sphere in a $d$-dimensional spacetime and leave an Euclidean space equipped with a reduced metric, the remaining spacetime will have the topology of $\mathbb{R}^2$ with the Euler characteristic, $\chi=1$. Although the dimensional truncation in this way is destructive to the topology of a whole spacetime, i.e., it radically changes the curvature of a whole spacetime, the reduced metric in the temporal and radial parts still keeps all of the global information and properties in the $d$-dimensional spacetime, which will be shown in the Hawking temperature. 

Based on the Euler characteristic $\chi$ of compact spaces with boundaries, a topological formula was introduced~\cite{Robson:2018con} for temperature in the natural units,
\begin{eqnarray}
	T_{\text{topology}}=\frac{1}{4\pi \chi}\Sigma_{j\leq\chi}\int_{r_{H_j}}{\sqrt{g}\,\mathcal{R}\,dr},
	\label{temperature}
\end{eqnarray}
where the Ricci scalar $\mathcal{R}$ is the function of only radial coordinate $r$, $g$ is the Euclidean metric determinant, and the symbol $\Sigma_{j\leq\chi}$ denotes the summation over the location of $j$-th Killing horizon $r_{H_j}$. 
When applying this topological formula in the truncated Gauss-Bonnet black hole with the Euler characteristic, $\chi=1$, we find that the temperature is exactly the Hawking temperature, Eq.~\eqref{eq:temp}, i.e., $T_{\text{topology}}=T_{\rm H}$.

From the above result, we can clearly observe that the Hawking temperature is a purely topological quantity. Indeed, the indications that topological signals and structures do show up in black hole thermodynamics have been provided~\cite{Yerra:2022coh,Wei:2021vdx,Zhang:2020kaq,Ovgun:2019ygw} in some specific models. A recent study in the Unruh-DeWitt model shows~\cite{Tjoa:2022oxv} that the two-dimensional truncation allows us to extract many physical results that agree with those in the four-dimensional spacetime. For spherically symmetric black holes, the two-dimensional description also suggests that the information of horizon temperature is stored in the truncated spacetime.

\subsection{Maxwell's equal area law and coexistence curves}
\label{sec:Maxwells}
Before discussing the Maxwell equal area law, we write the state equation from Eq.~\eqref{eq:temp},
\begin{eqnarray}
	P=\frac{q^2 e^{-{k}/{r_+^2}}+3 r_+^3 \left[r_+ \left(2 \pi  r_+ T-1\right)+8 \pi  \alpha  T\right]}{8 \pi  r_+^6}.\label{eq:steq}
\end{eqnarray}
The critical point is located at the inflection point of critical isotherm, and is determined by
	\begin{equation}\label{eq:staph}
	\left(\frac{\partial P}{\partial
		r_+}\right)_T=0=\left(\frac{\partial^2P}{\partial r^2_+}\right)_T.
\end{equation}
 For the branch of analytical solutions, Eq.~\eqref{eq:staph} has the following two positive real roots,\footnote{For the large nonlinear electrodynamic parameter, $k\gg 1$, the critical value is consistent~\cite{Hyun:2019gfz} with that of $q=0$.} 
\begin{align}
	r_c&=2\sqrt{3\alpha}, \qquad q\rightarrow0, \nonumber\\
	r_c&=(5q^2)^{1/4}, \qquad \alpha, k\rightarrow0.
\end{align}
We then obtain the critical thermodynamic quantities,
\begin{eqnarray}
	T_c=\frac{q^2 e^{-{k}/{r_c^2}} \left(k-3 r_c^2\right)+3 r_c^6}{3 \pi  r_c^5 \left(12 \alpha +r_c^2\right)},
\end{eqnarray}
\begin{eqnarray}
	P_c=\frac{e^{-{k}/{r_c^2}} \left\{q^2 \left[2 k \left(4 \alpha +r_c^2\right)-12 \alpha  r_c^2-5 r_c^4\right]+3 r_c^6\, e^{{k}/{r_c^2}} \left(r_c^2-4 \alpha \right)\right\}}{8 \pi  r_c^8 \left(12 \alpha +r_c^2\right)}.
\end{eqnarray}

The coexistence curve of small and large black holes can be given when we analyze the characteristic swallow tail behavior of the Gibbs free energy. It is usually governed by the Clausius-Clapeyron equation,
\begin{eqnarray}
	\frac{dP}{dT} =\frac{S_l-S_s}{V_l-V_s},
\end{eqnarray} 
where $V_{s(l)}$ and $S_{s(l)}$ denote the thermodynamic volume and entropy for small (large) black holes, respectively. Alternatively, the coexistence curve can be determined when we construct the Maxwell equal area law. The coexisting phases have the same Gibbs free energy, so we have 
\begin{eqnarray}
		-SdT+\Phi dQ+VdP=0.\label{eq:dG0}
\end{eqnarray} 
Integrating Eq.~\eqref{eq:dG0} at constant $T, Q$, we find the equal area law on the $(P,V)$ plane,
\begin{eqnarray}
	\oint VdP=0.\label{eq:equarea}
\end{eqnarray} 
Along the coexistence curve,  the two states have the same pressure.
Using Eq.~\eqref{eq:equarea}, we have the following two equations,
\begin{align}
	\label{eq:eqlaw}
	&P(r_s,T)=P(r_l,T)=P^*, \nonumber\\
	&P^*\cdot(V_l-V_s)=\int_{r_s}^{r_l}P(r_+,T)dV,
\end{align}
where $P^*$ denotes an isobar, $r_s$ and $r_l$ denote the horizon radii of small and large black holes, respectively. Substituting Eq.~\eqref{eq:steq} into Eq.\eqref{eq:eqlaw}, we give Maxwell's equal area law,
\begin{align}
	\label{eq:eqlaw1}
	\frac{q^2 r_s^5 e^{-{k}/{r_l^2}}}{r_l}-\frac{q^2 r_l^5 e^{-{k}/{r_s^2}}}{r_s}+3 r_l^2 r_s^2 \left\{8 \pi  \alpha  \left(r_s^3-r_l^3\right) T+r_sr_l\left(r_s-r_l\right)  \left[r_s \left(2 \pi  r_l T-1\right)-r_l\right]\right\}=0,   \notag\\ 
	\left(r_s^4-r_l^4\right) \left\{3 \left(r_s-r_l\right) r_s^3 r_l^3 e^{k ({1}/{r_l^2}+{1}/{r_s^2})} \left\{8 \pi  \alpha  \left(r_s^2+r_l r_s+r_l^2\right) T 
	 + r_s r_l \left[r_s \left(2 \pi  r_l T-1\right)-r_l\right]\right\} \right.\notag\\ \left.
	  + q^2 r_s^6 e^{{k}/{r_s^2}}-q^2 r_l^6 e^{{k}/{r_l^2}}\right\}=0.
\end{align}

\begin{figure}[h]
	\begin{center}
		\includegraphics{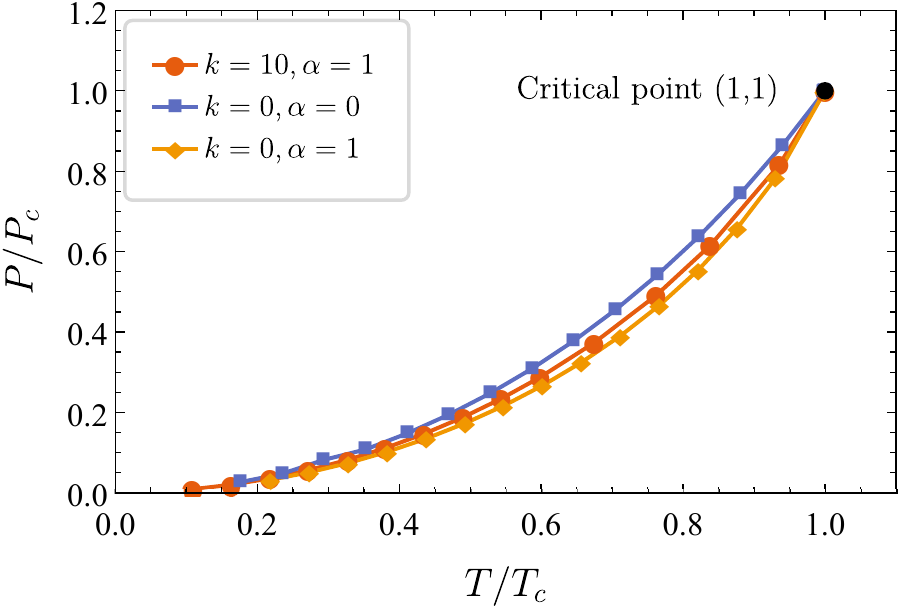}
	\end{center}
	\caption{Coexistence curves on the $({T}/{T_c}, {P}/{P_c})$ plane for small and large charged AdS black holes in the Gauss-Bonnet gravity coupled with nonlinear electrodynamics.}
	\label{fig:coexi}
\end{figure}

 We solve Eq.~\eqref{eq:eqlaw1} numerically for the case of $q=1$.  The numerical results of Maxwell's equal area law  are shown in Table \ref{tab:root}. We can clearly see that the radii of small and large black holes, $r_s$ and $r_l$, converge to the critical value, $r_c=3.464102$, for the case of  parameters $k=10$ and $\alpha=1$, when the temperature approaches to its critical value, $T_c=0.045944$, and in this situation the pressure also stays at its critical value, $P^*=P_c=0.003316$. For the other case of  parameters $k=0$ and $\alpha=1$, when the temperature goes to its critical value, $T_c=0.045641$, $r_s$ and $r_l$ equal their critical radius, $r_c=3.552759$, and the pressure also reaches the critical value, $P^*=P_c=0.003251$. When the temperature is falling, the radius of small black holes $r_s$ decreases, while the radius of large black holes $r_l$ increases. Moreover, we can check that the small and large black holes share the same Gibbs free energy. From Maxwell's equal area law, we can determine the coexistence curve on the $(P, T)$ plane, as shown in Fig.~\ref{fig:coexi}. It is worth to note that these coexistence curves with different values of $k$ and $\alpha$ exhibit the same behavior as that of those related to the van der Waals fluid~\cite{Kubiznak:2012wp}, i.e., all the coexistence
curves have a positive slope everywhere and terminate at the critical point. We find that the small/large black hole transition occurs in the Gauss-Bonnet gravity coupled with nonlinear electromagnetic field when $T<T_c$.   Furthermore, we note that a large nonlinear electrodynamic parameter $k$ leads to a slight increase of pressure for a fixed temperature, which can be observed from the comparison between the orange curve $(k=10, \alpha=1)$ and the yellow curve $(k=0,\alpha=1)$; however, a large Gauss-Bonnet coupling parameter $\alpha$ leads to a decrease of pressure for a fixed temperature, which can be observed from the comparison between the yellow curve $(k=0,\alpha=1)$ and  the blue curve $(k=0, \alpha=0)$ in Fig.~\ref{fig:coexi}.

\begin{table}[h]
	\centering
		\begin{tabular}{ cccc}
				\hline
					\hline
			\multicolumn{4}{c}{$k=10, \alpha=1$}\\
			\hline
			$T$ & $r_s$ & $r_l$ &  $P^*$   \\
			\hline
			0.045944 & 3.464102 & 3.464102 & 0.003316  \\
			0.043000  & 1.902183  & 5.004910& 0.002707  \\
			0.040000  & 1.520203 & 5.936279& 0.002240 \\
			0.038510  &1.388779 & 6.386753&0.002039  \\
			0.035000   & 1.149546 & 7.493093& 0.001627 \\
			0.032510& 1.015734 & 8.354845& 0.001376  \\
			0.031000 & 0.943953 & 8.923261& 0.001237 \\
			0.027500  & 0.796818 & 10.42368 & 0.000952\\
			\hline
			\multicolumn{4}{c}{$k=0, \alpha=1$}\\
			\hline
			0.045641 & 3.552759 & 3.552759 & 0.003251  \\
			0.042500 & 2.028396 & 6.420123 & 0.002551  \\
			0.040000 & 1.727614 & 7.762185 & 0.002141 \\
			0.037500 & 1.535324 & 9.038833 & 0.001803  \\
			0.035000 & 1.396224 & 10.34403 & 0.001517 \\
			0.032500 & 1.288937 & 11.73540 & 0.001271  \\
			0.030000 & 1.202839 & 13.26463 & 0.001056 \\
			0.027500 & 1.131836 & 14.98951 & 0.000869\\
				\hline
				\hline
		\end{tabular}
	\caption{The numerical results of equations governed by the Maxwell equal area law.}\label{tab:root}
\end{table}

\subsection{Critical exponents}
\label{sec:Exponent}
Critical exponents describe~\cite{Wei:2014qwa,Miao:2016ieh} the behavior of physical quantities in the vicinity of critical points. Following Ref.~\cite{Kubiznak:2012wp}, we denote the dimensionless quantities around a critical point as follows,
	\begin{equation}\label{eq:dimenless}
	p=\frac{P}{P_c}, \qquad t=\frac{T}{T_c}-1, \qquad  \omega=\frac{v}{v_c}-1,
\end{equation}
where $v=(4/3)r_+$ is the specific volume. The critical exponents $\alpha$, $\beta$, $\gamma$, and $\delta$ are defined by
\begin{align}
	&C_v= T\frac{\partial S}{\partial T}\Big|_v\propto|t|^{-\alpha},\nonumber\\
	&\eta= v_l-v_s\propto|t|^\beta,\nonumber\\
    &\kappa_T= -\frac{1}{v}\frac{\partial v}{\partial P}\Big|_T\propto|t|^{-\gamma},\nonumber\\
	&|P-P_c|\Big|_{T_c}\propto|v-v_c|^\delta.\label{defsofcriexpo}
\end{align}
Note that $t$ and $\omega$ are infinitesimal when expansions are made near the critical point.
Using Eq.~\eqref{eq:dimenless}, we can expand the state equation, Eq.~\eqref{eq:steq}, in the following form,
\begin{equation}\label{eq:steq1}
	p=1+At+Bt\omega+C\omega^3+\mathcal{O}(t\omega^2,\omega^4),
\end{equation}
where the coefficients read 
\begin{eqnarray}
	A&=&\frac{3T_c(4\alpha+r^2_c)}{4r^3_c},\nonumber\\
	B&=&-\frac{3 T_c \left(12 \alpha +r_c^2\right)}{4P_c r_c^3},\nonumber\\
	C&=&\frac{1}{P_c}\left[ -\frac{q^2 e^{-{k}/{r_c^2}} \left(27 k^2 r_c^2-96 k r_c^4+84 r_c^6-2 k^3\right)}{12 \pi  r_c^{12}}-\frac{3 T_c \left(40 \alpha +r_c^2\right)}{4 r_c^3}+\frac{3}{2 \pi  r_c^2}\right].
\end{eqnarray}

Because the small/large black hole transition occurs under the condition, $T<T_c$, i.e., $t<0$, we compute the critical exponents in this situation.  Differentiating  Eq.~\eqref{eq:steq1} with respect to $\omega$, we have
\begin{equation}
dp=(Bt+3C\omega^2)d\omega.\label{diffeqofsta}
\end{equation}
Considering Maxwell's equal area law in the vicinity of critical points, we obtain the following two equations from Eq.~(\ref{eq:steq1}), Eq.~(\ref{eq:equarea}) and Eq.~(\ref{diffeqofsta}),
\begin{align}
   &p=1+At+Bt\omega_l+C\omega^3_l=1+At+Bt\omega_s+C\omega^3_s,\nonumber\\
	&\int_{\omega_l}^{\omega_s}\omega(Bt+3C\omega^2)d\omega=0,
\end{align}
which give rise to one unique non-trivial solution,
\begin{equation}
	\omega_l=-\omega_s=\sqrt{\frac{-Bt}{C}}.\label{omegalequals}
\end{equation}
Because the entropy, Eq.~\eqref{eq:entropy}, leads to a vanishing heat capacity at constant volume, we derive from Eq.~(\ref{defsofcriexpo}) the first critical exponent,
\begin{equation}
C_v=0 \quad\Rightarrow\quad  \alpha=0. 
\end{equation}
Then, by using Eq.~(\ref{defsofcriexpo}) together with Eq.~(\ref{omegalequals}), we get the second critical exponent,
\begin{eqnarray}
	\eta\equiv v_l-v_s=v_c(\omega_l-\omega_s)=2v_c\omega_l=2v_c\sqrt{\frac{-Bt}{C}}\propto\sqrt{t} \quad\Rightarrow\quad  \beta=\frac{1}{2}.
\end{eqnarray}
Next, differentiating Eq.~\eqref{eq:steq1} with respect to $P$ on the both sides and ignoring the term proportional to $\omega^2$, we calculate the third exponent $\gamma$ by considering Eq.~(\ref{defsofcriexpo}),
\begin{eqnarray}
	\kappa_T\equiv -\frac{1}{v}\frac{\partial v}{\partial P}\Big|_T\propto\frac{1}{Bt}\quad\Rightarrow\quad  \gamma=1.
\end{eqnarray}
At last, considering the critical isotherm, $t=0$, in Eq.~\eqref{eq:steq1} and  using Eq.~(\ref{defsofcriexpo}), we obtain the fourth exponent,
\begin{eqnarray}
	p-1=C\omega^3\quad\Rightarrow\quad  \delta=3.
\end{eqnarray}
We clearly see that the critical exponents associated with the charged AdS black holes in the Gauss-Bonnet gravity coupled with nonlinear electrodynamics coincide with those of the van der Waals fluid.

 \section{Thermodynamic cycles and heat engines}
 \label{sec:heat engine}
\begin{figure}[h]
	\begin{center}
		\includegraphics{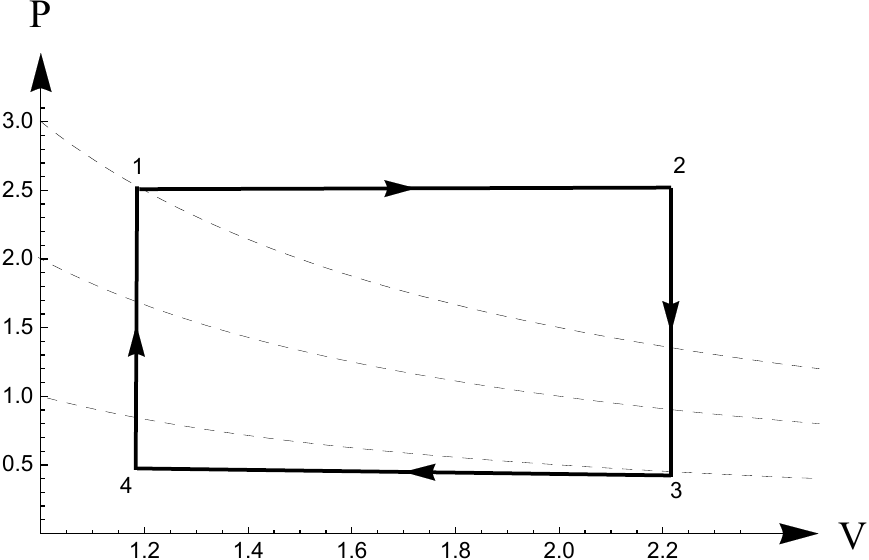}
	\end{center}
	\caption{Schematic diagram of a heat engine cycle.  The solid black rectangle denotes the heat engine cycle and the dashed curves stand for the $P-V$ diagram of this cycle.}
	\label{cycle}
\end{figure}

 \setcounter{equation}{0}
In the above section, we analyze the thermodynamic properties of charged AdS black holes in the Gauss-Bonnet gravity coupled with nonlinear electrodynamics, including the thermodynamic stability associated with nonlinear electrodynamics, derive the Hawking temperature from the perspective of topology,
investigate Maxwell's equal area law and coexistence curves, and compute the critical exponents of this kind of black holes.
These performances provide the basic help for our following study of heat energy. In this section, we propose a new kind of heat engines by regarding such a charged AdS black hole as a working substance. Following  Johnson's natural explanation~\cite{Johnson:2014yja} for the cyclic choice of static black holes, we consider an engine cycle composed of two isobars and two isochores, as shown in Fig.~\ref{cycle}. The work done by such an engine along our cycle can easily be calculated by the formula,
\begin{eqnarray}
	W=\oint PdV.\label{work}
\end{eqnarray} 

For most black holes a valid engine cycle suggests that the  net input heat flow $Q_{\rm in}$ occurs along the top isobar and the  net output heat flow $Q_{\rm out}$ along the bottom isobar. Furthermore, an explicit expression of $C_P$ has been derived in Eq.~\eqref{heatcap}, which allows us to give the net inflow of heat along the upper isobar,
\begin{equation}
	Q_{\rm in}=\int_{T_1}^{T_2} C_P(P_1, T)\,dT.\label{inputhf}
\end{equation}
Dividing the net output work $W$ by the net input heat flow $Q_{\rm in}$ and using Eqs.~(\ref{heatcap}), (\ref{work}), and (\ref{inputhf}),  we obtain the exact expression for the heat engine efficiency,
\begin{align}
\eta=\frac{W}{Q_{\rm in}}=4 (P_1-P_4)\left\{\frac{\pi ^{5/3} q^2}{k} \left[\exp \left(-\frac{\pi ^{4/3} k \left(\sqrt{64 \pi ^4 \alpha ^3+S_1^2}-S_1\right)^{2/3}}{\left[\left(\sqrt{64 \pi ^4 \alpha ^3+S_1^2}-S_1\right)^{2/3}-4 \pi ^{4/3} \alpha \right]^2}\right)-1\right]\right. \notag\\ \left.
-
\frac{3{\pi}^{1/3} \left[\left(\sqrt{64 \pi ^4 \alpha ^3+S_1^2}-S_1\right)^{2/3}-4 \pi ^{4/3} \alpha \right]^2}{\left(\sqrt{64 \pi ^4 \alpha ^3+S_1^2}-S_1\right)^{2/3}}-
\frac{8 P \left[\left(\sqrt{64 \pi ^4 \alpha ^3+S_1^2}-S_1\right)^{2/3}-4 \pi ^{4/3} \alpha \right]^4}{\left(\sqrt{64 \pi ^4 \alpha ^3+S_1^2}-S_1\right)^{4/3}}\right. \notag\\ \left.
+
\frac{8 P \left[\left(\sqrt{64 \pi ^4 \alpha ^3+S^2}-S_2\right)^{2/3}-4 \pi ^{4/3} \alpha \right]^4}{\left(\sqrt{64 \pi ^4 \alpha ^3+S_2^2}-S_2\right)^{4/3}}
+\frac{3 {\pi}^{1/3} \left[\left(\sqrt{64 \pi ^4 \alpha ^3+S_2^2}-S_2\right)^{2/3}-4 \pi ^{4/3} \alpha \right]^2}{\left(\sqrt{64 \pi ^4 \alpha ^3+S_2^2}-S_2\right)^{2/3}}\right. \notag\\ \left.
+
\frac{\pi ^{5/3} q^2}{k} \left[1-\exp \left(-\frac{\pi ^{4/3} k \left(\sqrt{64 \pi ^4 \alpha ^3+S_2^2}-S_2\right)^{2/3}}{\left[\left(\sqrt{64 \pi ^4 \alpha ^3+S_2^2}-S_2\right)^{2/3}-4 \pi ^{4/3} \alpha \right]^2}\right)\right]\right\}^{-1} \notag\\
\times
\left\{\frac{\left[\left(\sqrt{64 \pi ^4 \alpha ^3+S_2^2}-S_2\right)^{2/3}-4 \pi ^{4/3} \alpha \right]^4}{\left(\sqrt{64 \pi ^4 \alpha ^3+S_2^2}-S_2\right)^{4/3}}-\frac{\left[\left(\sqrt{64 \pi ^4 \alpha ^3+S_1^2}-S_1\right)^{2/3}-4 \pi ^{4/3} \alpha \right]^4}{\left(\sqrt{64 \pi ^4 \alpha ^3+S_1^2}-S_1\right)^{4/3}}\right\},\label{heatengeff}
\end{align}
where the subscripts denote the quantities evaluated at the corners of the rectangle cycle labeled by $(1, 2, 3, 4)$ in Fig.~\ref{cycle}. 

\begin{figure}[h]
\begin{center}
   		\includegraphics[width=.5\linewidth]{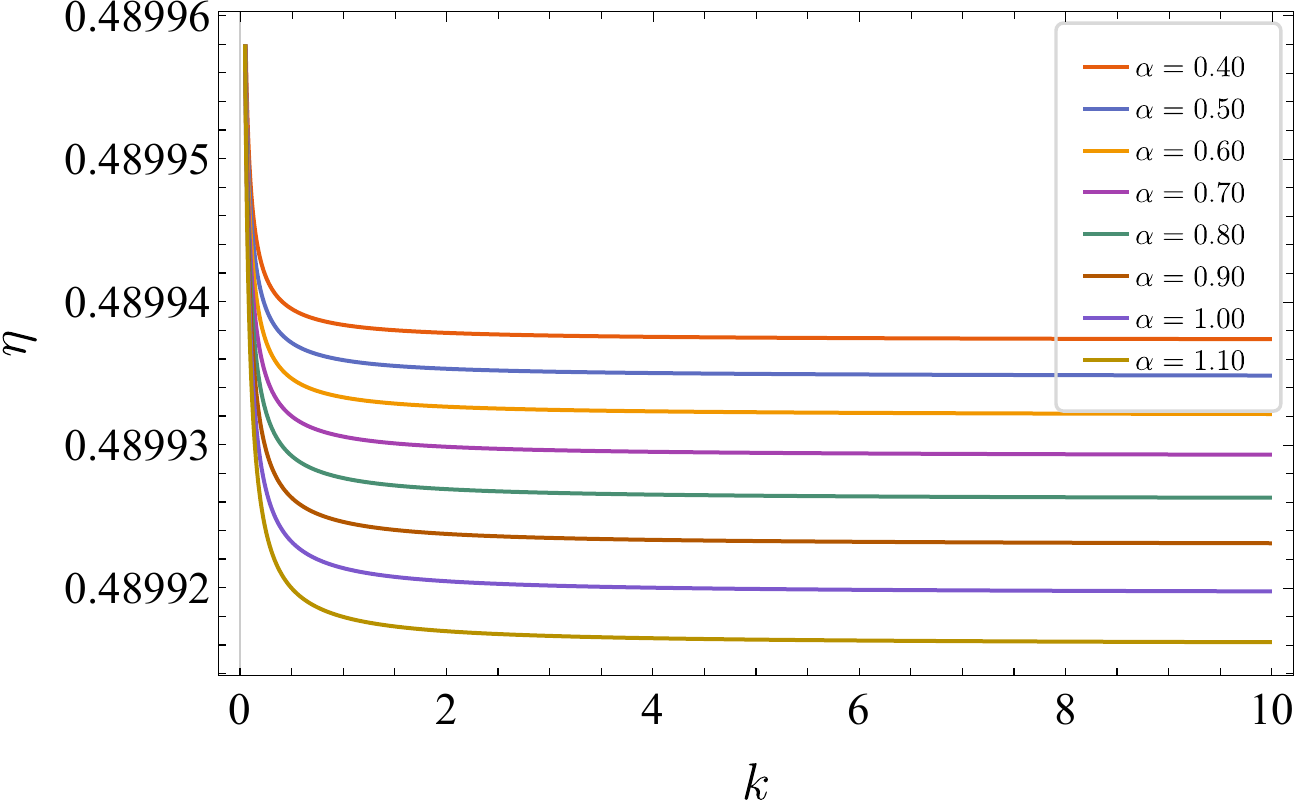}
   	\end{center}
   	\caption{The engine efficiency $\eta$ versus the nonlinear electrodynamic parameter $k$ for different values of $\alpha$. The set of sample values is taken as $P_1=50$, $P_4=1$, $S_1=1$, $S_2=500$, and $q=1$.}	\label{fig:nonlinear}
   \end{figure}

Recently, the holographic heat engines have been extensively discussed  \cite{Hennigar:2017apu,Nam:2019zyk,Balart:2019uok,Rajani:2019ovp,ElMoumni:2021zbp} in the theories of modified gravity.
Based on the results from the topological Einstein-Maxwell black hole, the Born-Infeld black hole,  the Kerr-AdS black hole, and the super-entropic black hole, a claim was made~\cite{Hennigar:2017apu} that the presence of nonlinear Born-Infeld  fields leads to an increase of efficiency. However, we have an opposite observation for the charged AdS black hole in the Gauss-Bonnet gravity coupled to nonlinear electrodynamics, that is, we observe a significant decrease of efficiency as the nonlinear electrodynamic parameter $k$ increases, and present such a behavior by plotting the lengthy Eq.~\eqref{heatengeff} in Fig.~\ref{fig:nonlinear}.
   
\begin{figure}[ht]
	\begin{subfigure}{.5\textwidth}
		\centering
		\includegraphics[width=.9\linewidth]{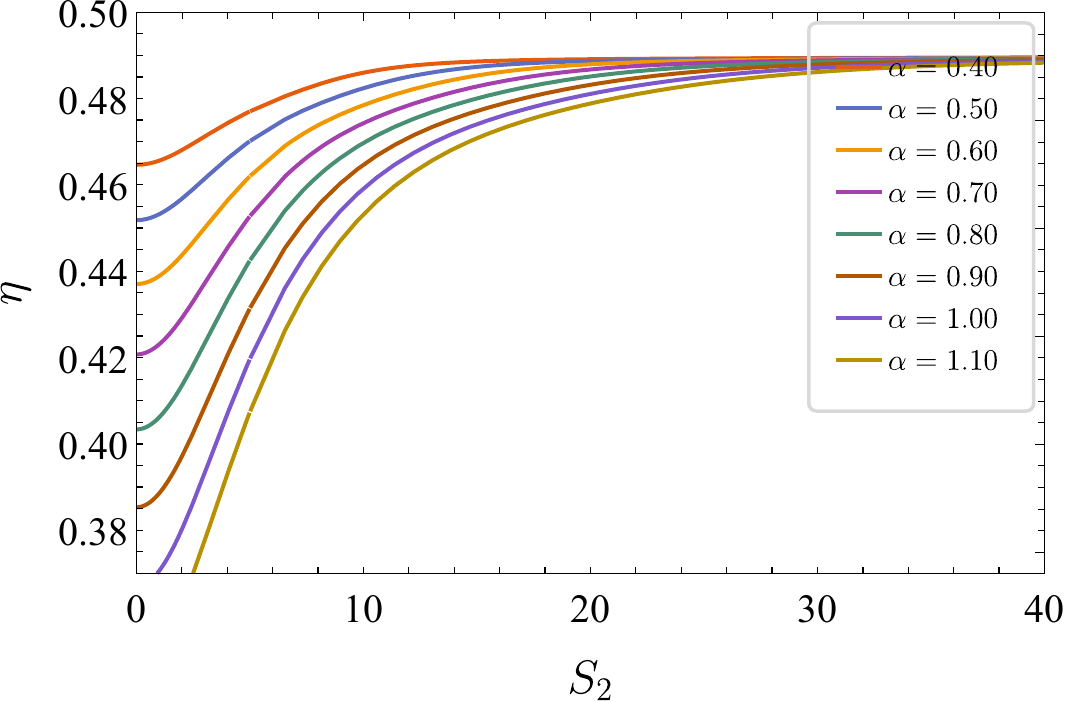}  
		\caption{The engine efficiency $\eta$}
		\label{fig:ratio1}
	\end{subfigure}
	\begin{subfigure}{.5\textwidth}
		\centering
		\includegraphics[width=.9\linewidth]{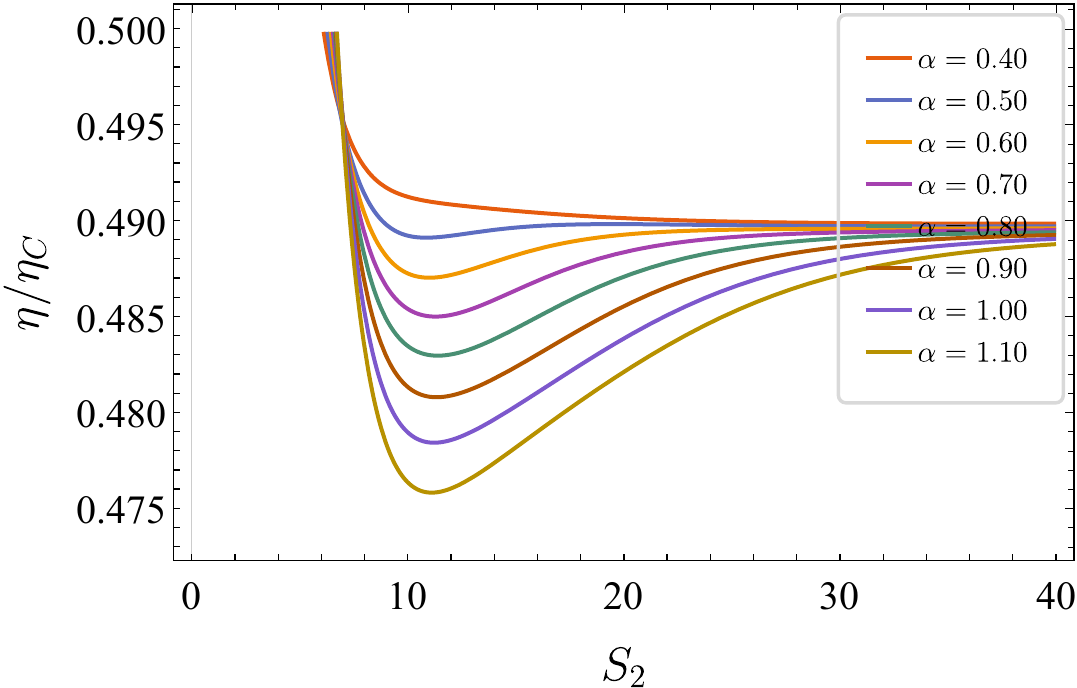}  
		\caption{Ratio of engine efficiency $\eta/\eta_{\rm C}$}
		\label{fig:ratio2}
	\end{subfigure}
	\caption{(a) The engine efficiency $\eta$ versus  entropy $S_2$ and (b) the ratio $\eta/\eta_{\rm C}$ versus  entropy $S_2$ for different values of $\alpha$. The set of sample values is taken as $P_1=50$, $P_4=1$, $S_1=5$, $k=0.5$, and $q=1$.}
	\label{fig:ratio}
\end{figure}

Exact properties of heat engines depend on their working substances. Efficiency can reflect the intrinsic properties of a black hole as a working substance. And the efficiency of an engine is bounded by the Carnot efficiency determined by the temperature difference of two heat reservoirs,
   \begin{eqnarray}
  	\eta_{\rm C}=1-\frac{T_{\rm cold}}{T_{\rm hot}},
  \end{eqnarray}
 where $T_{\rm cold}$ and $T_{\rm hot}$ are the absolute temperatures of cold and hot reservoirs, respectively. This is the maximum efficiency that any heat engines can reach when operating between the same cold and hot reservoirs. In order to investigate the engine efficiency provided by the charged AdS black hole in the Gauss-Bonnet gravity coupled to nonlinear electrodynamics as a working substance, for a set of sample values we plot the corresponding efficiency $\eta$ and the ratio of $\eta$ to the Carnot efficiency $\eta_{\rm C}$.  On the left, diagram (a) of Fig.~\ref{fig:ratio}, the efficiency slowly grows as entropy $S_2$ increases and it approaches to the maximum efficiency, $\eta_{\rm max}\sim0.5$. On the right, diagram (b) of Fig.~\ref{fig:ratio}, there exists one point of inflection from which the efficiency ratio $\eta/\eta_{\rm C}$ increases as the entropy decreases. The reason is that the Carnot heat engine efficiency is going to zero when the two isotherm corners are approaching to each other.

\begin{figure}[h]
 \begin{center}
  		\includegraphics[width=0.48\linewidth]{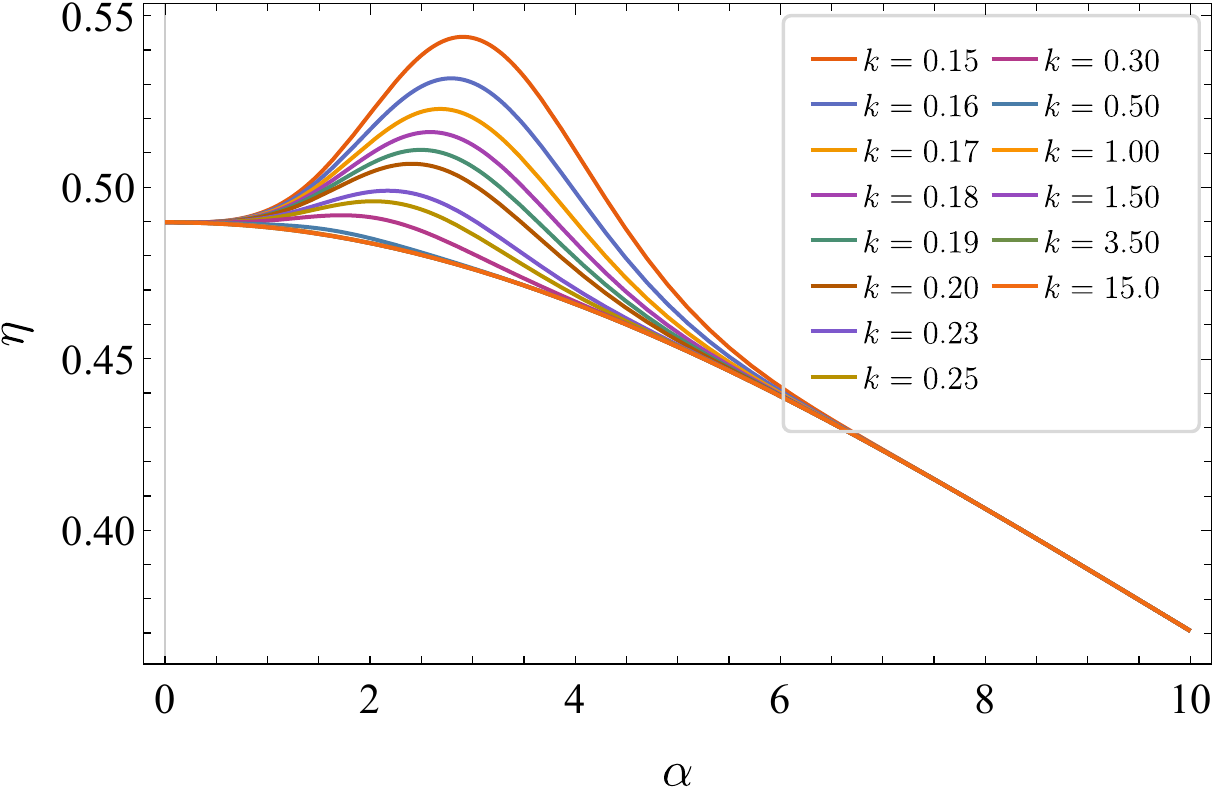}
  	\end{center}
  	\caption{The engine efficiency $\eta$ versus the Gauss-Bonnet coupling parameter $\alpha$ for different values of $k$. The set of sample values is taken as $P_1=50$, $P_4=1$, $S_1=1$, $S_2=50$, and $q=1$.}\label{fig:alp}
  \end{figure}

 We also highlight the higher curvature effect on the efficiency of the rectangular engine cycle in Fig.~\ref{fig:alp}. The effect of higher curvatures depends on the Gauss-Bonnet coupling parameter in general. We analyze the engine efficiency with respect to parameter $\alpha$ and find that the Gauss-Bonnet coupling leads to a decrease of efficiency  for spherical black holes.  This result is consistent with that observed~\cite{Johnson:2015ekr} in the Gauss–Bonnet–Einstein–Maxwell gravity. We emphasize that both the effect of nonlinear electrodynamics and that of Gauss-Bonnet couplings reduce the efficiency of the engine cycle depicted in Fig.~\ref{cycle}, see Fig.~\ref{fig:nonlinear} and Fig.~\ref{fig:alp}.  In other words, the nonlinear electrodynamic parameter and the Gauss-Bonnet coupling parameter have the similar effect on the efficiency reduction because they affect the black hole thermodynamics and the ability for a black hole to do mechanical work.
 When the parameter $\alpha$ is small ($\alpha$ is taken between $1$ and $5$ in Fig.~\ref{fig:alp}), the effect of nonlinear electrodynamic parameter $k$ is obvious, i.e., the efficiency is gradually increasing to its maximum at first and then decreasing as $k$ becomes large. But the curves with different values of $k$ are going to coincide when the Gauss-Bonnet coupling parameter $\alpha$ is large ($\alpha>6$ in Fig.~\ref{fig:alp}), indicating that the effect of nonlinear electrodynamics will be attenuated in the strong Gauss-Bonnet coupling region.
     
   \section{Conclusions and discussions } 
   \label{sec:con}
   
We investigate the nonlinear electrodynamic effect on the thermodynamic stability, the topological property of Hawking radiations, and the holographic heat engine in the framework of the Gauss-Bonnet gravity.
We make a general analysis of black hole thermodynamics based on heat capacity, and show that a thermodynamically unstable region appears for a negative heat capacity, but such a region disappears due to an increase of the nonlinear electrodynamic parameter.
In addition, we reproduce the Hawking temperature in the truncated spacetime using a topological approach. Our result indicates that the Hawking temperature is a topological quantity and its information is stored in the truncated spacetime. 
Even though the above results are obtained in the five-dimensional model given in Sec.~\ref{sec:thermodynamics}, we confirm that they are applicable in any dimensional spacetime at least for static and spherically symmetric black holes.

We discuss Maxwell's equal area law and coexistence curves and find the existence of a van der Waals-like phase transition, in particular, we give the reason for the van der Waals-like phase transition based on critical exponents.
We also propose a holographic heat engine in which a charged AdS black hole in the  Gauss-Bonnet gravity coupled with nonlinear electrodynamics is regarded as a working substance, and derive the analytic expression of efficiency for such an engine. We observe that the nonlinear electrodynamic effect leads to a decrease of efficiency of the rectangular engine cycle, and so does the effect from the Gauss-Bonnet coupling. Further, the nonlinear electrodynamic effect will be attenuated with an increase of the Gauss-Bonnet coupling parameter. We emphasize that such an observation is opposite to that already given in literature for some models of modified gravity.

For the charged AdS black holes in the Gauss-Bonnet gravity coupled with nonlinear electrodynamics, we reveal new thermodynamic properties and provide a specific approach under which the constructed holographic heat engine can reach high efficiency. In the former aspect, we find the critical behavior via the small/large black hole phase transition, which is similar to that of van der Waals fluid. In particular, for the definition of critical exponents, we take the specific volume instead of the thermodynamic volume used in the previous literature~[60, 47]. In the latter aspect, we suggest a novel holographic heat engine in which the charged AdS black hole acts as a working material and investigate its efficiency with respect to the two parameters, the nonlinear electrodynamic parameter $k$ and the Gauss-Bonnet coupling parameter $\alpha$. In particular, we analyze the relationship between the efficiency of such an engine and the parameter spaces of $\alpha$ and $k$. Our results may give some insights into the fundamental relationship between the Gauss-Bonnet gravity theory and the nonlinear electrodynamics.

 \paragraph{Acknowledgments}
 The authors would like to thank the anonymous referee for the helpful comments that
 improve this work greatly.
 This work was supported in part by the National Natural Science Foundation of China under Grant Nos. 11675081 and 12175108. 
      
    \appendix 

\section{Derivation of Eq.~\eqref{eq:eqlaw1}}
\label{sec:maxwell}
\setcounter{equation}{0}
Substituting Eq.~\eqref{eq:steq} into Eq.~\eqref{eq:eqlaw}, we get the following equations,
\begin{equation}
	\label{eq:pstar1}
	\frac{q^2 e^{-k/r_s^2}+3 r_s^3 \left[r_s \left(2 \pi  r_s T-1\right)+8 \pi  \alpha  T\right]}{8 \pi  r_s^6}=P^*,
\end{equation}
\begin{equation}
		\label{eq:pstar2}
	\frac{q^2 e^{-k/r_l^2}+3 r_l^3 \left[r_l \left(2 \pi  r_l T-1\right)+8 \pi  \alpha  T\right]}{8 \pi  r_l^6}=P^*,
\end{equation}
\begin{eqnarray}
	\label{eq:eqlaw3}
	\frac{P^*\pi^2}{2}  \left(r_l^4-r_s^4\right)=\int_{r_s}^{r_l} \frac{q^2 e^{-k/r_+^2}+3 r_+^3 \left[r_+ \left(2 \pi  r_+ T-1\right)+8 \pi  \alpha  T\right]}{8 \pi  r_+^6} \, d\left(\frac{\pi ^2}{2}  r_+^4\right).
\end{eqnarray}
By eliminating the parameter $P^*$ in Eq.~\eqref{eq:pstar1} and Eq.~\eqref{eq:pstar2}, we derive the first equation of Eq.~\eqref{eq:eqlaw1}.
Then, integrating Eq.~\eqref{eq:eqlaw3} after substituting Eq.~\eqref{eq:pstar1} into it, we obtain 
      \begin{eqnarray}
      	\label{eq:eqlaw4}
    \frac{r_s^6r_l^6}{8k}\left\{q^2 \left(e^{-k/r_s^2}-e^{-k/r_l^2}\right)+k \left(r_s-r_l\right) \left[T \left(48 \pi  \alpha +4 \pi  r_s^2+4 \pi  r_l r_s+4 \pi  r_l^2\right)-3 r_s-3 r_l\right]\right\} \nonumber\\
    + \frac{r_l^6}{16}	\left(r_l^4-r_s^4\right) \left\{q^2 e^{-k/r_s^2}+3 r_s^3 \left[r_s \left(2 \pi  r_s T-1\right)+8 \pi  \alpha  T\right]\right\} =0.
      \end{eqnarray}
  Similarly, substituting Eq.~\eqref{eq:pstar2} into Eq.~\eqref{eq:eqlaw3} yields
  \begin{eqnarray}
  	\label{eq:eqlaw5}
  \frac{r_s^6r_l^6}{8k}	\left\{q^2 \left(e^{-k/r_s^2}-e^{-k/r_l^2}\right)+k \left(r_s-r_l\right) \left[T \left(48 \pi  \alpha +4 \pi  r_s^2+4 \pi  r_l r_s+4 \pi  r_l^2\right)-3 r_s-3 r_l\right]\right\} \nonumber\\
  	+\frac{r_s^6}{16}\left(r_l^4-r_s^4\right) \left\{q^2 e^{-k/r_l^2}+3 r_l^3 \left[r_l \left(2 \pi  r_l T-1\right)+8 \pi  \alpha  T\right]\right\}=0.
  \end{eqnarray}
  Comparing Eq.~\eqref{eq:eqlaw4} with Eq.~\eqref{eq:eqlaw5}, we finally reach the second equation of Eq.~\eqref{eq:eqlaw1}.

\bibliographystyle{JHEP}
\bibliography{references}

\end{document}